\documentclass[prl,nofootinbib]{revtex4}
\usepackage{graphicx}
\usepackage{latexsym}
\def\be{\begin{equation}}
\def\ee{\end{equation}}
\def\bea{\begin{eqnarray}}
\def\eea{\end{eqnarray}}

\begin{document}
\title{Probing Inflation and Dark Energy with Current Cosmological Observations}

\author{Jun-Qing Xia${}^a$}
\email{xiajq@mail.ihep.ac.cn}

\author{Gong-Bo Zhao${}^a$}
\email{zhaogb@mail.ihep.ac.cn}

\author{Bo Feng${}^b$}
\email{fengbo@resceu.s.u-tokyo.ac.jp}

\author{Xinmin Zhang${}^a$}
\email{xmzhang@mail.ihep.ac.cn}

\affiliation{${}^a$Institute of High Energy Physics, Chinese
Academy of Science, P.O. Box 918-4, Beijing 100049, P. R. China}

\affiliation{ ${}^b$ Research Center for the Early
Universe(RESCEU), Graduate School of Science, The University of
Tokyo, Tokyo 113-0033, Japan}

\date{\today}

\begin{abstract}

It is commonly believed that our Universe has experienced at least
two different stages of accelerated expansion. The early stage is
known as inflation and the current acceleration is driven by dark
energy. Observing inflation and dark energy dynamics are among the
most important aspects of the current cosmological research. In this
paper we make a first detailed probe of the possible degeneracies
between dynamical inflation and dark energy in light of the current
cosmological observations. We have combined Type-Ia supernova (Riess
"gold" samples), galaxy clustering (SDSS 3D power spectra) and the
Wilkinson Microwave Anisotropy Probe (WMAP) observations and
performed a global analysis using the Markov chain Monte Carlo
(MCMC) method. We find the inclusion of inflation and dark energy
parameters together make the parameter spaces broader and the
degeneracy between the inflation dark energy parameters are
nontrivial: the allowed/preferred behavior of dynamics in one sector
is dependent on the prior of the other sector. Interestingly a
deviation from scale-invariant primordial spectrum is sightly more
preferred by the current cosmological observations when one
marginalizes over dynamical dark energy models.

~

PACS number(s): 98.80.Es, 98.80.Cq

\end{abstract}

\maketitle

\section{I. Introduction}
%{\it Introduction. --- }
Recent advances in both theoretical and observational cosmology have
revealed that our Universe has experienced two different stages of
accelerated expansion. One is the inflation in the very early
universe when its tiny patch was superluminally stretched to become
our observable Universe today\cite{inflation,Guth:1979bh}. It can
naturally explain why the universe is flat, homogeneous and
isotropic. Inflation is driven by a potential energy of a scalar
field called inflaton and its quantum fluctuations turn out to be
the primordial density fluctuations which seed the observed
large-scale structures (LSS) and anisotropy of cosmic microwave
background radiation (CMB). In the past decade, inflation theory has
successfully passed several nontrivial tests. In particular, the
released first year Wilkinson Microwave Anisotropy Probe (WMAP) data
\cite{kogut} have detected a large-angle anti-correlation in the
temperature--polarization cross-power spectrum, which is the
signature of adiabatic superhorizon fluctuations at the time of
decoupling\cite{WMAP2}. The cosmological observations such as the
first year WMAP(WMAP1) and the Sloan Digital Sky Survey (SDSS) are
in good agreement with adiabatic and scale invariant primordial
spectrum(see e.g. \cite{0310723,Seljak:2004xh}), which is consistent
with single field slow rolling inflation predictions. On the other
hand, however, inflation generically predicts primordial spectrum
with some deviations from scale invariance and it is crucial to
probe the scale dependence of primordial spectrum in light of the
cosmological observations. The tentatively interesting features of
the first year WMAP observations have aroused a lot of interests to
explain such effects with the dynamics of
inflation\cite{Contaldi:2003zv}.

%All these observational data are in good agreement with the basic
%picture of the inflation paradigm and the idea of inflation has now
%been proved in this sense. These data, however, have also revealed
%many interesting detailed features that cannot be explained in the
%simplest version of the inflation model, hence we are making much
%effort to construct a realistic particle physics model of inflation.

In 1998 the analysis of the redshift-distance relation of Type Ia
supernova (SNIa) has revealed the existence of the second stage of
accelerated expansion that has started rather recently when a
mysterious new energy component dubbed dark energy(DE) has dominated
the energy density of the Universe.~\cite{Riess98,Perl99}. Recent
observations of SNIa have confirmed the accelerated expansion at
high confidence level\cite{Tonry03,Riess04,Riess05,Astier:2005qq}.
The nature of dark energy is among the biggest problems in modern
physics and has been studied widely. A cosmological constant, the
simplest candidate of DE where the equation of state (EOS) $w$
remains -1, suffers from the well-known fine-tuning and coincidence
problems. Alternatively, dynamical dark energy models with the
rolling scalar fields have been proposed, such as
quintessence\cite{quint,pquint}, phantom~\cite{phantom} and
k-essence\cite{Chiba:1999ka,kessence}. Given that currently we know
very little on the theoretical aspects of dark energy, the
cosmological observations play a crucial role in our understanding
of dark energy. If the observations strongly favor the dark energy
component with a deviation from $w=-1$, then the cosmological
constant which puzzles the theorists would not be the source driving
current acceleration of the universe.

Since there are many common aspects of the background physics,
albeit the striking difference of the energy scale between the two
stages of accelerated expansion, it would be nice if one could unify
inflation with dark energy. The model of quintessential
inflation\cite{Pvilenkin} has been proposed in history trying to
unify the two epoches of accelerated expansions. The first year WMAP
observations show the lack of the temperature-temperature
correlation power on the largest scales and many studies have
attributed it to the suppressed primordial spectrum due to some
dynamics of inflaton. Intriguingly, in Ref.\cite{Moroi:2003pq} the
authors have attributed the lack of power to the isocurvature
fluctuations in the quintessence which may be generated during
inflation, thus the different dynamics of the dark energy sector and
inflation can lead to similar effects on the observations, where
such a degeneracy may reflect the possible connections between dark
energy and inflation.

Observing inflation and dark energy dynamics are among the most
important aspects of the current cosmological research. So far to
our knowledge in the previous observational studies of dynamical
dark energy or inflation not enough attention has been paid to the
possible correlation between the two sectors: dark energy and
inflation. In this paper we make a first detailed probe of the
possible degeneracies between dynamical inflation and dark energy in
light of the current cosmological observations. Our paper is
structured as follows: in Section II we describe the method and the
data; in Section III we present our results by global fittings;
finally we give our summary and conclusions in Section IV.

\section{II. Method and data}

The role of the primordial scalar spectral index $n_s$ in the
probe of inflation is somewhat similar to that of dark equation of
state $w$ in the measurement of dark energy. The case of $n_{s}=1$
corresponds to the scale invariant Harrison-Zel'dovich spectrum,
where the scalar spectrum is constant over all scales and $w=-1$
corresponds to the cosmological constant where the density of dark
energy does not evolve with time. A detection of deviation from
$n_{s}=1$ or $w=-1$ would mark a breakthrough in our understanding
of inflation and dark energy. To make our study on inflation and
dark energy symmetric and in some sense model independent, we
parametrize $n_s$ and $w$ with linear expansions and keep them to
the first order:
\begin{equation}
\label{ns} n_s(k)=n_s(k_{s0}) + \alpha_{s} \ln \left(
\frac{k}{k_{s0}}\right),
\end{equation}
where $k_{s0}$ is a pivot scale which is arbitrary in principle and
$\alpha_{s}$ is a constant characterizing the "running" of the
scalar spectral index. The scalar spectral index  $n_s$ is related
to the primordial scalar power spectrum ${\cal P}_{\chi}(k)$  by
definition:
\begin{equation}
\label{nsdef} n_s(k) \equiv d{\cal P}_{\chi}(k)/d \ln k +1 .
\end{equation}
Correspondingly ${\cal P}_{\chi}(k)$ is now parametrized
as\cite{paraPk}:
\begin{equation}
\label{spectrum} \ln {\cal P}_{\chi}(k)=\ln A_{s} +
(n_{s}(k_{s0})-1)\ln \left(
\frac{k}{k_{s0}}\right)+\frac{\alpha_{s}}{2}\left(\ln
\left(\frac{k}{k_{s0}}\right)\right)^{2} .
\end{equation}
Accordingly for the dark energy sector the equation of state is
parametrized as \cite{Linderpara}:
\begin{equation}
\label{EOS} w_{DE}(a) = w_{0} + w_{1}(1-a)
\end{equation}
where $a=1/(1+z)$ is the scale factor and $ w_{1}$ characterizes the
"running" of the equation of state. The equation of state is defined
as the pressure over the energy density, which leads to the
evolution of dark energy density via the energy conservation of dark
energy:
\begin{equation}
\rho_Q(a)/\rho_Q(a_0=1)=a^{-3(1+w_0+w_1)} \exp{[3w_1(a-1)]}~.
\end{equation}

The current publicly available codes like
CMBfast\cite{cmbfast,IEcmbfast} and CAMB\cite{camb,IEcamb} have
well adapted the constant running of the primordial spectral
index. For the constant running of the dark energy equation of
state it is not so straightforward\cite{Vikman,Perturbation}. For
the common dynamical dark energy models like quintessence or
phantom, the equation of state does not get across -1, and for the
global fittings through parametrizations of $w$ the dynamics of
quintessence or phantom has been fully enclosed in $w$ in which
sense one can even reconstruct the potentials of quintessence and
phantom with $w$. However in the cases where $w$ gets across -1
during evolution, single field scalar dark energy models like
quintessence, phantom or k-essence cannot realize such a
transition and on the other hand, for the parametrizatized dark
energy one is encountered with the problem of divergence for the
perturbation equations\cite{Perturbation}. These properties have
been investigated in details in our companion paper in
Ref.\cite{Perturbation}. Dark energy with equation of state
getting across -1 has been dubbed quintom in the sense that it
resembles the combined behavior of quintessence and
phantom\cite{fengcross,Feng:2004ff,Guo:2004fq}. We have shown that
in general to realize the model of quintom one needs to add extra
degrees of freedom beyond the standard single field
case\cite{Perturbation}. In the extant two-field-quintom model and
the single field model with a high derivative
term\cite{Li:2005fm}, the perturbation of dark energy is shown to
be continuous when $w$ gets across -1.

For the parametrization of the equation of state which gets across
-1, we introduce a small positive constant $\epsilon$ to divide the
full range of the allowed value of  $w$ into three parts: 1) $ w
> -1 + \epsilon$; 2) $-1 + \epsilon \geq w  \geq-1 - \epsilon$; and 3) $w < -1 -\epsilon $.
Working in the conformal Newtonian gauge, one can easily describe
the perturbations of dark energy as follows \cite{ma}: \bea
    \dot\delta&=&-(1+w)(\theta-3\dot{\Phi})
    -3\mathcal{H}(c_{s}^2-w)\delta~~, \label{dotdelta}\\
\dot\theta&=&-\mathcal{H}(1-3w)\theta-\frac{\dot{w}}{1+w}\theta
    +k^{2}(\frac{c_{s}^2\delta}{{1+w}}+ \Psi)~~ . \label{dottheta}
\eea

Neglecting  the entropy perturbation contributions, for the regions
1) and 3) the equation of state does not get across $-1$  and
perturbations are well defined by solving
Eqs.(\ref{dotdelta},\ref{dottheta}). For the case 2), the
perturbation of energy density $\delta$ and divergence of velocity,
$\theta$, and the derivatives of $\delta$ and $\theta$ are finite
and continuous for the realistic quintom dark energy models. However
for the perturbations of the parametrized quintom there is clearly a
divergence. In our study for such a regime, we match the
perturbation in region 2) to the regions 1) and 3) at the boundary
and set\cite{Perturbation}
\begin{equation}\label{dotx}
  \dot{\delta}=0 ~~,~~\dot{\theta}=0 .
\end{equation}
In our numerical calculations we've limited the range to be $|\Delta
w = \epsilon |<10^{-5}$ and we find our method is a very good
approximation to the multi-field quintom. For more details of this
method we refer the readers to our previous companion paper
\cite{Perturbation}.

In this study we have implemented the publicly available Markov
Chain Monte Carlo package CosmoMC\cite{CosmoMC}, which has been
modified to allow for the inclusion of Dark Energy perturbation with
the equation of state getting across $-1$. We assume purely
adiabatic initial conditions and a flat Universe. Our most general
parameter space is:
\begin{equation}
\label{parameter} {\bf P} \equiv (\omega_{b}, \omega_{c},
\Theta_{s}, \tau, w_{0}, w_{1}, n_{s}, \alpha_{s},
\ln(10^{10}A_{s}),r)
\end{equation}
where $\omega_{b}\equiv\Omega_{b}h^{2}$ and
$\omega_{c}\equiv\Omega_{c}h^{2}$ are the physical baryon and Cold
Dark Matter densities relative to the critical density, $\Theta_{s}$
is the ratio (multiplied by 100) of the sound horizon to the angular
diameter distance at decoupling, $\tau$ is the optical depth to
re-ionization, $A_{s}$, $n_{s}$ and $\alpha_{s}$ characterize the
primordial scalar power spectrum. $r$ is the tensor to scalar ratio
of the primordial power spectrum and for the tensor slope $n_T$ we
have fixed $r=-8 n_T$. For the pivot of the primordial spectrum we
set $k_{s0}=0.05$Mpc$^{-1}$. Furthermore, we make use of the Hubble
Space Telescope (HST) measurement of the Hubble parameter
$H_{0}\equiv 100$h~km~s$^{-1}$~Mpc$^{-1}$\cite{Hubble} by
multiplying the likelihood by a Gaussian likelihood function
centered around $h=0.72$ and with a standard deviation
$\sigma=0.08$. We also impose a weak Gaussian prior on the baryon
and density $\Omega_{b}h^{2}=0.022\pm0.002$ (1 $\sigma$) from Big
Bang Nucleosynthesis\cite{BBN}.

In our calculations we have taken the total likelihood to be the
products of the separate likelihoods (${\bf \cal{L}}_i$) of CMB, LSS
and SNIa. In other words defining $\chi_{i}^2 \equiv -2 \log {\bf
\cal{L}}_i$, we get \be\label{chi2} \chi^2_{total} = \chi^2_{CMB} +
\chi^2_{LSS} + \chi^2_{SNIa} ~ . \ee We have included the first-year
temperature and polarization data with the routine for computing the
likelihood supplied by the WMAP team\cite{WMAPdadax,kogut}. We use
the code developed in Ref.\cite{SDSS} to fit the 3D power spectrum
of galaxies from the SDSS and the bias factor of SDSS has been used
as a continuous parameter to give the minimum $\chi^2$ value. For
the SNIa data we use the 157 gold-sample data published by Riess
\emph{et al}\cite{Riess04} and in the calculation of the likelihood
we have marginalized over the nuisance parameter\cite{SNcode}.

Compared with Riess "gold" sample of SNIa, the first year Supernova
Legacy Survey (SNLS) sample\cite{Astier:2005qq} has been measured
with one telescope and the systematics are better constrained.
However for the time being their determination on cosmological
parameters are marginal comparable\cite{globalxia}. Due to the fact
that the two samples of SNIa have been from different methods of
light curve fitting a combination of them is not so straightforward.
For LSS for the time being the Two degree Field (2dF)
\cite{Cole:2005sx} and SDSS\cite{SDSS} are also comparable and for
the current study we have not combined the two data sets
simultaneously.

For each regular calculation, we run 8 independent chains comprising
of $150,000-300,000$ chain elements and spend thousands of CPU hours
to calculate on a supercomputer. The average acceptance rate is
about $40\%$. We test the convergence of the chains by Gelman and
Rubin criteria\cite{R-1} and find $R-1$ is of order $0.01$ which is
more conservative than the recommended value $R-1<0.1$.

\section{III.  Results}

To get the clear effects on the possible correlations between
dynamical inflation and dark energy we mainly focus on the three
kinds of models: one is the $\Lambda$CDM cosmology where $n_s,r$ and
$\alpha_s$ are enclosed in the inflationary sectors, one is with
scale-invariant primordial spectrum (r not included) where $w_0$ and
$w_1$ are included in the dark energy sector, and the third one
includes all of the free parameters like $n_s, r, \alpha_s, w_0$ and
$w_1$ \footnote{For the simplicity of discussions we do not discuss
the effects on $A_s$ and $\Omega_{DE}$, although we did vary them in
our fittings.}.

\begin{table*}
TABLE 1. Median values and 1$\sigma$ constrains on $n_s, r,
\alpha_s, w_0$ and $w_1$ for the three models as explained in the
text, shown together with the minimum $\chi^2$ values . For $r$ the
2$\sigma$ upper bounds have been shown instead.
\begin{center}
\begin{tabular}{|c|ccc|}
\hline

&$\Lambda$CDM &Scale Invariant&Dynamical DE\\

\hline

$w_0$  &set to
-1&$-1.28\pm0.23$ &$-1.28\pm0.25$ \\

$w_1$    &set to 0&$0.554^{+0.880}_{-0.948}$
&$0.593^{+0.815}_{-0.899}$  \\

$n_s$   &$0.990\pm0.041$ &set to 1 &$0.984\pm0.057$ \\

$\alpha_s$   &$-0.0322^{+0.0357}_{-0.0353}$ &set to 0
&$-0.0400^{+0.0366}_{-0.0355}$   \\

$r$          &$<0.455(95\%)$ &set to 0     &$<0.465(95\%)$   \\

  \hline

$\chi^2$ &1638&1635&1634\\

\hline
\end{tabular}
\end{center}

\end{table*}

Table 1 lists all of the relevant 1-dimensional median values and
1$\sigma$ constrains for the three models specified above. Shown
together are the minimum $\chi^2$ values for each case. For the
constraints on $r$ only 2$\sigma$ upper bounds have been shown. The
effects of different priors on the inflationary and dark energy
sectors will be explained in details with the 2-dimensional graphics
of the sub parameter space below.

\begin{figure}[htbp]
\begin{center}
\includegraphics[scale=0.7]{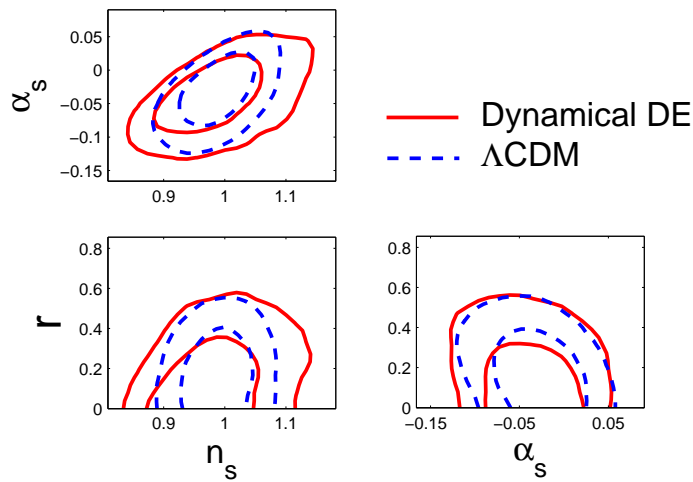}
\caption{ Constraints on the primordial spectrum parameters by
marginalizing over $\Lambda$CDM cosmology (Blue Dash line) and over
dynamical dark energy with equation of state $w= w_0+ w_1 (1-a)$
(Red Solid line).  The two dimensional contours stand for $68\%$ and
$95\%$ C.L. \label{fig1}}
\end{center}
\end{figure}

In Fig. \ref{fig1} we delineate the two dimensional contours of the
primordial spectrum parameters by marginalizing over $\Lambda$CDM
cosmology (Blue Dash lines) and over dynamical dark energy with
equation of state in Eq. (\ref{EOS}) (Red Solid lines). The contours
stand for $68\%$ and $95\%$ C.L. Although in general the case with
dynamical dark energy gives less stringent constraints on the
primordial spectrum, the $\Lambda$CDM case does not overlap fully
with the dynamical dark energy case. Intriguingly for the
combination with the inclusion of dark energy dynamics, the
constraints on the tensor contributions exhibit different behaviors
at $68\%$ and $95\%$ C.L. And at $68\%$ C.L. the constraint on $r$
is more stringent with the inclusion of dark energy dynamics, which
is different from naively expected. This can be explained by the
mild preference on the dark energy dynamics in the combination of
WMAP1+SDSS+SNIa(Riess) and the degeneracy between dark energy
dynamics and $r$ on the largest scales of CMB. From Table 1 we can
find the preference of dynamical DE is slightly larger than
1$\sigma$ from the different minimum $\chi^2$ values as well as from
the 1-dimensional constraint on the parameter of $w_0$.
 Also we can see some
difference in the constraints on $\alpha_s$ and the value of $n_s$
at 0.05 Mpc$^{-1}$. Note the best fit values of $\alpha_s$ in some
sense deviate a lot from zero, the difference on the probe of scale
variance may be non-negligible for the two cases. We get $\alpha_s =
-0.032^{+0.046}_{-0.035}$ for the $\Lambda$CDM cosmology and
$-0.040^{+0.037}_{-0.036}$ instead for the case with dynamical dark
energy. Roughly speaking a scale invariant primordial spectrum is
not yet ruled out in the $\Lambda$CDM cosmology while disfavored
more than $1 \sigma$ when we allow the dynamics of dark energy! We
will turn to this in more details in the later part of this paper.

\begin{figure}[htbp]
\begin{center}
\includegraphics[scale=0.4]{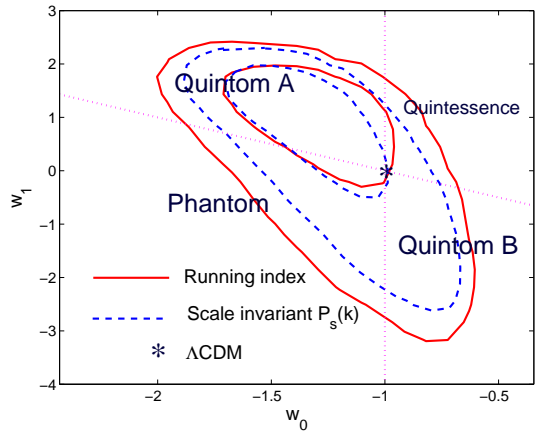}
\caption{$68\%$ and $95\%$ constraints on the dynamical dark energy
model $w= w_0+ w_1 (1-a)$ marginalizing over scale-invariant
primordial spectrum (Blue Dash line) and over primordial spectrum
with a running spectral index (Red Solid line). In the latter case
tensor contributions are included. The Dotted lines stand for
$w_{0}\equiv-1$ and $w_{0}+w_{1}\equiv-1$. \label{fig2}}
\end{center}
\end{figure}

In Fig. \ref{fig2} we plot the $68\%$ and $95\%$ constraints on the
$(w_0,w_1)$ contour marginalizing over scale-invariant primordial
spectrum (Blue Dash line) and over primordial spectrum with a
running spectral index (Red Solid line). We find interestingly the 1
$\sigma$ regions do not overlap fully and the cosmological constant
case lies at the edge where the 1 $\sigma$ regions
overlap.Numerically we get $w_0=-1.28\pm {0.23},
w_1=0.569^{+0.865}_{-0.927}$ for the scale-invariant case and
$w_0=-1.28\pm {0.25}, w_1=0.591^{+0.804}_{-0.881}$ with a running
spectral index.  We cut the parameter space of $w_{0}-w_{1}$ into
four areas by the line of $w_{0}\equiv-1$ and $w_{0}+w_{1}\equiv-1$,
Quintessence, Phantom, Quintom A ($w_{DE}$ is Phantom-like today but
Quintessence-like in the past) and Quintom B (Dark Energy has
$w_{DE}>-1$ today but $w_{DE}<-1$ at higher redshifts). Quintom A
almost occupies the 1 $\sigma$ region. In a previous companion paper
in the probe of dynamical dark energy, we got similar results where
for the primordial spectrum part we included only $n_s$ and
$A_s$\cite{globalxia}. However inflation predicts nonzero tensor
contribution and it is a crucial parameter towards verifying the
theory of inflation, in the current paper we consider $r$
contributions in cases we do not simply assume a Harrison-Zel'dovich
spectrum.

\begin{figure}[htbp]
\begin{center}
\includegraphics[scale=0.9]{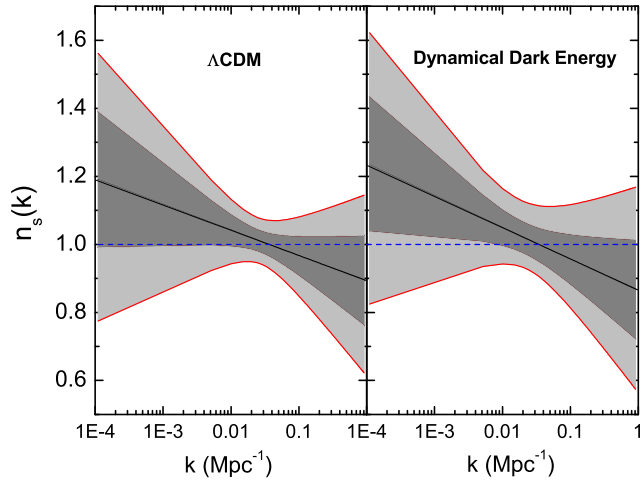}
\caption{$68\%$ and $95\%$ constraints on the slope of $n_s(k)$ by
marginalizing over $\Lambda$CDM cosmology (Left) and over dynamical
dark energy with equation of state $w= w_0+ w_1 (1-a)$ (Right). The
blue thick line stands for the scale-invariant primordial spectrum.
\label{fig3}}
\end{center}
\end{figure}

\begin{figure}[htbp]
\begin{center}
\includegraphics[scale=1.0]{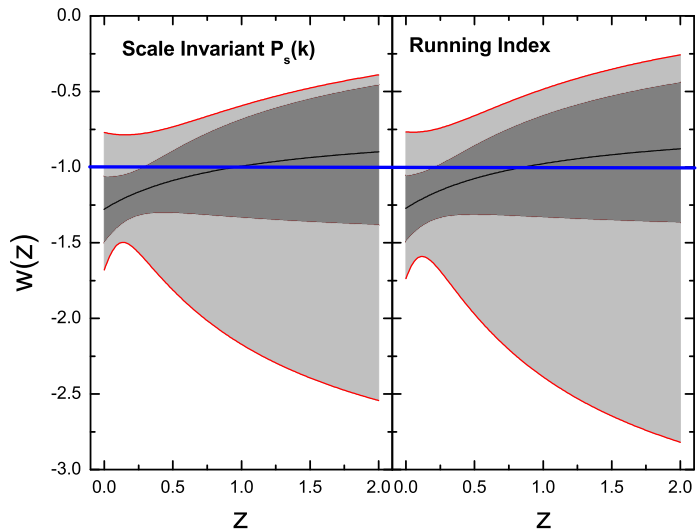}
\caption{$68\%$ and $95\%$ constraints on the evolution of dynamical
dark energy model $w= w_0+ w_1 (1-a)$ marginalizing over
scale-invariant primordial spectrum (Left) and over primordial
spectrum with a running spectral index (Right). In the latter case
tensor contributions are included. \label{fig4}}
\end{center}
\end{figure}

In the probe of deviations from scale invariance and dynamical dark
energy it is sometimes more intuitive to get constraints on the
whole slopes of $n_s(k)$ and $w(z)$ on the relevant scales. We have
got the constraints on $n_s(k)$ and $w(z)$ by using the covariance
matrix of $n_s(k_{s0}), \alpha_s$ and $w_0,w_1$ from our MCMC
results. In Fig. \ref{fig3} and Fig. \ref{fig4} we plot the
constraints on $n_s(k)$ and $w(z)$ for the two different cases. In
Fig. \ref{fig3} it is obvious that scale-invariant primordial
spectrum is more disfavored in the case with dynamical dark energy,
although it does allow more parameter space compared with the
$\Lambda$CDM cosmology. It is intriguing that if we start always
from $\Lambda$CDM cosmology in the probe of the primordial spectrum
information, we might fail to find the true physics which is crucial
for understanding inflation. Neglecting dark energy dynamics might
lead to the bias on the probe of inflation, although the statistical
significance is rather low.

In  Fig. \ref{fig4} we find that in both cases dynamical dark energy
is favored more than $1 \sigma$ compared with the cosmological
constant. This happens that in both cases $w_0<-1$ is favored more
than $1 \sigma$, as shown previously. The $1 \sigma$ regions in Fig.
\ref{fig4} differ little when marginalizing over scale-invariant
primordial spectrum and over primordial spectrum with a running
spectral index, which can also be easily understood from Fig.
\ref{fig2}\footnote{Ref.\cite{Seljak:2004xh} has done some similar
analysis on this point, where different observational datasets have
been used with somewhat different results. They did not include the
effects of dark energy perturbations and study the constraints on
dynamical dark energy with also $\alpha$ and $r$. In this paper we
have tried to probe the deviation from scale variance and dynamical
dark energy in equal weights.} and the minimum $\chi^2$ values in
Table 1. Our results here are somewhat similar to a previous
analysis in Ref. \cite{Corasaniti:2004sz}. The mild preference of
dynamical dark energy over a cosmological constant lies mainly on
the Riess sample of SNIa\cite{Riess04}. On the other hand such a
preference also shows imprints on large scales of CMB multipoles via
the Integrated Sachs-Wolfe (ISW) effect and dark energy
perturbations, which are correlated with $r$ and $\alpha_s$. Future
improvements on the SNIa observations will also inversely help to
probe the information of the primordial spectrum, as implied in our
Fig. \ref{fig1} and Fig. \ref{fig3}.

%{\it Conclusion. --- }
\section{IV. Summary and Discussions}

 In this paper we have preformed an analysis
of global fitting allowing simultaneously the dynamics in both
inflation and dark energy sector. Our result shows that there is
generically some correlation between the inflation and dark energy
parameters and the parameter space is typically enlarged for the
probe of dynamics in either sector compared to cases where one
assume there can be some dynamics in only one of these sectors. When
we allow some dynamics in the dark energy sector we find
interestingly a deviation from scale-invariant primordial spectrum
is sightly more favored by current cosmological observations than
assuming a $\Lambda$CDM cosmology.

  We should stress that starting from the theoretical aspects it is
not so natural to predict exactly a constant running on $\alpha_s$
or $w_1$, on the other hand given the current precision of the
observations and also for a generic study of the dynamics we are
simply using some of the parametrizations on each sector. In general
it is not so easy to achieve a precision cosmology
\cite{Bridle:2003yz}, given the fact that for CMB (also LSS and
lensing) on large scales we have the cosmic variance, on small
scales we have the secondary effects which are non-negligible
compared with the damped power itself and for LSS on small scales we
have the not yet well-understood factor of bias and the complicated
nonlinear evolutions. And for SNIa the calibration is in some sense
an open issue.  Actually with our better understandings of the
Universe and the developments of technology we are able to probe the
Universe with more versatile observations with time going on.
Nowadays we are also hoping to extract the information of the
Universe through the observations from weak and strong lensing,
radio galaxies, Gamma-ray bursts(GRB), Lyman-$\alpha$ forest(Lya)
and the bayonic oscillations inherited in the LSS observations.
However we also need to understand better their inherited
systematics before we can extract more robust and precise
information of our Universe. In this paper we have not considered
small scale CMB observations or those like Lya or GRB. We should
point out that our fittings are only one of the data combinations
and different combinations can certainly lead to somewhat different
fitting results, on the other hand given the data set we use our
results are robust.

Observing inflation and dark energy dynamics are among the most
important aspects of the current cosmological research. Currently we
are not yet able to detect the dynamics on either sector and in
particular the possible contaminations of the CMB observations on
the largest scales will affect our results
qualitatively\cite{Schwarz04}. However our main concentration is to
make a first detailed study in the probe of possible correlations
between inflation and dark energy. If one starts always from the
concordance $\Lambda$CDM cosmology one cannot achieve more subtle
physics beyond that. And as we have shown in the cases one assume
there is dynamics in only one of the sectors this can lead to some
bias in our probe of the dynamics. If such result can be confirmed
at great significance by future observations, it will also shed
light on the study of inherent relationship between inflation and
dark energy.

Recently the WMAP three year data (WMAP3) has  been
released\cite{Spergel:2006hy,Page:2006hz,Hinshaw:2006,Jarosik:2006,WMAP3IE},
with significant improvements on the quality\cite{Page:2006hz}.
WMAP3 team claims a nontrivial preference of the red-tilted power
spectrum, which is different from the first year results and we
can hopefully get some different fitting results. However our
qualitative results depend strongly on the largest scales of CMB
and as implied recently in Ref.\cite{Feng:2006ui} the pixel-based
likelihood of WMAP3 on the largest scales seems to take too large
a weight on the determination of cosmological parameters and one
might take the risk of getting biased, in the present paper we do
not include the new WMAP3 data for a reanalysis and leave it for
future investigations.

 {\it
Acknowledgement. --- } We acknowledge using the sources of Shanghai
Supercomputer Center (SSC) and  the Super Computing Center of
Chinese Academy of Sciences (SCCAS). We thank Hong Li, Mingzhe Li
and Jun'ichi Yokoyama for helpful discussions. This work is
supported in part by the National Natural Science Foundation of
China (NSFC) under Grant Nos. 10533010, 90303004, 19925523 and by
the Ministry of Science and Technology of China under Grant No.
NKBRSF G19990754.

\end{document}